\documentclass[a4paper]{aa}
\usepackage{ffbib, epsfig}
\usepackage{times}

\begin{document}
\thesaurus{6(08.09.2 EV Lac; 08.12.1; 08.01.2; 08.03.5; 13.25.5)}

\title{An extreme X-ray flare observed on EV~Lac by ASCA in July 1998}

\author{F.~Favata\inst{1} \and F.~Reale\inst{2} \and G.~Micela\inst{3}
  \and S.~Sciortino\inst{3} \and A.~Maggio\inst{3} \and
  H.~Matsumoto\inst{4}}

\institute{Astrophysics Division -- Space Science Department of ESA, ESTEC,
  Postbus 299, NL-2200 AG Noordwijk, The Netherlands
\and
Dip.\ Scienze FF.\ \& AA., Sez. Astronomia, Univ.\ Palermo,
Piazza del Parlamento 1, I-90134 Palermo, Italy
\and
Osservatorio Astronomico di Palermo, 
Piazza del Parlamento 1, I-90134 Palermo, Italy 
\and
Center for Space Research -- Massachusetts Institute of Technology,
77 Massachusetts Avenue, 02139 Cambridge (MA), USA
}

\offprints{F. Favata (ffavata@astro.estec.esa.nl)}

\date{Received Jun 6, 1999 ; accepted Sept. 28, 1999}


\maketitle 
\begin{abstract}
  
  We present a long (150~ks elapsed time) X-ray observation of the
  dM3.5e star EV~Lac, performed with the ASCA observatory in July
  1998, during which an exceptionally intense flaring event (lasting
  approximately 12~ks) was observed: at the flare's peak, the X-ray
  count rate in the ASCA GIS detectors was $\simeq 300$ times the
  quiescent value.  The physical parameters of the flaring region have
  been derived by analyzing the decay, using both a ``classic''
  quasi-static approach and an approach based on hydrodynamic
  simulations of decaying flaring loops.  Notwithstanding the large
  peak X-ray luminosity, this second method shows that the flare's
  decay is compatible with its being produced in a relatively compact
  region of semi-length $L \simeq 1.3 \times 10^{10}$ cm ($\simeq
  0.5~R_*$), large but not exceptional even by solar standards.  The
  flare decays is fast (with a measured $e$-folding time for the light
  curve of $\le 2$~ks), but nevertheless the hydrodynamic-based
  analysis shows strong evidence for sustained heating, with the shape
  of the light curve dominated by the time evolution of the heating
  rather than by the natural cooling of the flaring plasma. As a
  consequence, the quasi-static method yields a much larger estimate
  of the loop's length ($L \simeq 2~R_*$). The event shows (similarly
  to some other well-studied large stellar flares) a light curve
  characterized by two separate decay time constants (with the initial
  decay being faster) and a significant enhancement in the plasma
  metal abundance at the peak of the flare.  The energetics of the
  event are exceptional, with the peak X-ray luminosity of the event
  reaching up to $\simeq 25$\% of the bolometric luminosity of the
  star, making this one of the largest X-ray flare (in relative terms)
  observed to date on a main-sequence star.

  \keywords{Stars: individual: EV~Lac -- Stars: late-type -- Stars:
    activity -- Stars: coronae; X-rays: stars}

\end{abstract}

\section{Introduction}
\label{sec:intro}

One of the basic standing problems of stellar coronal astronomy is the
determination of the spatial structuring of the coronal plasma.  While
for the study of stellar interior structure the first-order
approximation of spherical symmetry is a good starting point, stellar
coronae are, as shown by the large body of extant imaging observations
of the solar corona, far from spherically symmetric. The solar corona
shows a high degree of spatial structuring: most of the X-ray luminous
plasma is confined in coronal loops preferentially located at
mid-latitudes, with an average position which tracks the migration of
sunspots through the solar cycle. The lack of spatial information
constitutes a strong limitation for the study of stellar coronae:
low-resolution coronal X-ray spectra are insensitive to the plasma's
density, so that non-dispersive, CCD- or proportional counter-based
spectral observations do not allow to distinguish between a large
diffuse corona and a compact, structured, high-pressure one. It is
thus not possible, if the solar analogy is postulated, to determine
how the solar corona scales toward higher levels of activity, i.e.\ if
largely through a spatial filling of the available volume with coronal
loops (yielding a relatively symmetric corona) or if through the
filling of a relatively small number of coronal structures with
significantly higher density plasma.

Thus far the main tools to study the spatial distributions of the
coronal plasma have been eclipse experiments and the study of flares.
While the observation of rotational modulation should also in
principle allow to derive the spatial distribution of the emitting
plasma, as discussed by \cite*{sch98} convincing examples of
rotationally modulated X-ray emission are rare. Under a given set of
assumptions, the study of the decay phase of a flare can yield
information about the size of the flaring structure; different methods
for this type of analysis have been developed and applied in the past
to several observations of stellar flares. The widely applied
quasi-static approach (see below) almost invariably results, when
applied to intense stellar flares, in long, tenuous coronal loops
extending out to several stellar radii. The stronger flares yield in
general larger sizes.  Detailed hydrodynamic modeling of flaring loops
has provided useful insight on stellar X-ray flares (\cite{rps+88});
more recently, diagnostic tools have been developed for the derivation
of the size of stellar flaring loops and of the heating evolution
(\cite{rsp93}; \cite{rbp+97}; \cite{rm98}).  In the solar case, in
addition to the ``compact'' flares, in which the plasma appears to be
confined to a single loop whose geometry does not significantly evolve
during the event (an assumption shared by the quasi-static method and
by the hydrodynamic modeling quoted above), a second class of flares
is usually recognized, i.e.\ the ``two-ribbon'' events, in which an
entire arcade of loops first opens and then closes back; the
footpoints of the growing system of loops are anchored in
H$\alpha$-bright ``ribbons''.  These flares are generally
characterized by a slower rise and decay, and a larger energy release.
Compact flares have often been considered to be due to ``impulsive''
heating events, while the longer decay times of two-ribbon events have
been considered as a sign of sustained heating. However, also in the
case of compact flares sustained heating has been shown to be
frequently present (\cite{rbp+97}), so that the distinction may indeed
be less clear than often thought.

Fitting of X-ray spectra with physical models of static loops
(\cite{mp97}; \cite{smf+99}) can also yield the surface filling factor
of the plasma as one of the fit parameters. However, for loops smaller
than the pressure scale height this method only gives an upper limit
to the filling factor, which needs to be further constrained, for
example, with estimates of the plasma density based on EUV line ratios
(\cite{mp97}). With few exceptions\footnote{Notably the observation of
  $\alpha$~CrB, \cite*{sk93}, in which the observed total eclipse
  constrains the corona on the G5V star to have a scale height much
  less than a solar radius.}, eclipse studies of the quiescent
emission of eclipsing binaries have thus far failed to yield strong
constraints on spatial structuring of the plasma (\cite{sch98}).  In
part this is due to the inversion of the weak observed modulation
being mathematically an intrinsically ill-posed problem, where few
compact structures can mimic the emission from a more diffuse medium
(see discussion in \cite{sch98}). Recently, the observation of the
total eclipse of a large flare on Algol (\cite{sf99}) has for the
first time yielded a strong geometrical constraint on the size of a
flaring structure.  The geometrical loop size is significantly smaller
than the size derived from the analysis of the flare's decay
(\cite{fs99}) using the quasi-static method, showing how such approach
can over-estimate the actual loop size. The characteristics of the
Algol flare are such that the hydrodynamic decay, sustained heating
framework which we also use here yields a large range of allowed loop
semi-lengths, with the lower end of the range marginally compatible
with the geometrically derived size.

The presence of intense X-ray flares on flare stars\footnote{We use
  the term to mean ``M-type dwarfs, either single or members of a
  multiple system, which show frequent sudden enhancements of their
  X-ray, UV and optical luminosity''.} was well established with
\emph{Einstein} observations -- with some by now classic observations
such as the one relative to a flare on Proxima Cen (\cite{hai83}).
However, \emph{Einstein} observations were usually relatively short
(few ks) thus imposing a bias on the type of events which could be
detected (\cite{asg87}). In particular, the short total integration
times reduced the chance of detecting rare event types. The EXOSAT
observatory featured long, uninterrupted observations, and thus
allowed to collect a more unbiased view of flares on low-mass stars,
resulting in a database (\cite{pts90}) of about 300~hr of flare stars
observations, from which it is apparent that flares come in a large
variety of sizes for both their time scales and their energetics.
\cite*{pts90} did not attempt to derive the spatial scales of the
observed flares, although they tentatively divided them into two
classes reminiscent of the solar compact and two-ribbon flares.  The
next generation of of flare observations came with the ROSAT All-Sky
Survey (RASS), which, thanks to its sky scanning strategy, allowed to
search for rare, long-lasting events. Some flares of previously
undetected magnitude and duration are indeed present in it
(\cite{sch94}); in particular, EV~Lac showed a long X-ray flare
lasting approximately a day, superimposed on a much shorter but more
intense event.

We have observed EV~Lac for two days with the ASCA observatory,
detecting the most intense X-ray flare thus far observed on a
main-sequence star, with a 300-fold peak increase of the X-ray count
rate. This paper presents a detailed analysis of the flaring event,
and it is organized as follows: the ASCA observations and their
reduction, together with the determination of the spectral parameters
for the flare are discussed in Sect.~\ref{sec:analy}, the parameters
of the flaring region are determined (using both the quasi-static
formalism of \cite{om89} and the hydrodynamic decay, sustained heating
framework of \cite{rbp+97}) in Sect.~\ref{sec:flare}, with a
discussion (Sect.~\ref{sec:disc}) closing the paper.

\begin{figure*}[thbp]
  \begin{center}
    \leavevmode \epsfig{file=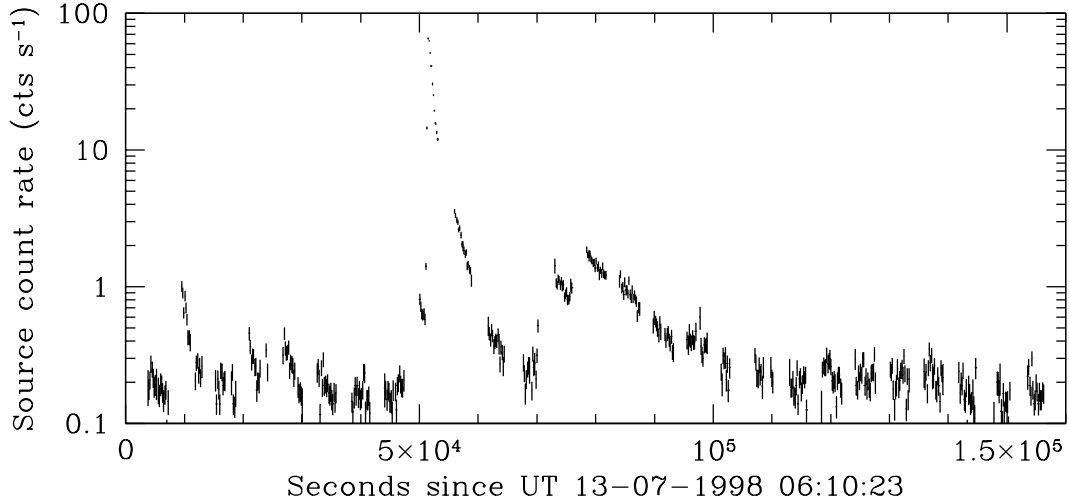, width=15.0cm, bbllx=20pt,
      bblly=400pt, bburx=600pt, bbury=700pt, clip=}
    \caption{The background-subtracted (background count rate $\simeq 0.05$
      cts s$^{-1}$) GIS-3 light curve of EV~Lac for the entire ASCA
      observation, binned in 180~s intervals.}
    \label{fig:lctotal}
  \end{center}
\end{figure*}

\section{Observations and data reduction}
\label{sec:analy}

EV~Lac was observed by ASCA continuously for $\simeq 150$~ks from 13
July 1998 06:10 UT to 15 July 1998 01:40 UT, as an AO-6 guest
investigator target (P.I.\ F.~Favata). The data were analyzed using
the {\sc ftools}~4.1 software, extracting both the spectra and the
light curves with the {\sc xselect} package and performing the
spectral analysis with the {\sc xspec}~10.0 package.  The {\sc mekal}
plasma emission model (\cite{mkl95}) was adopted for the spectral
analysis.  The peak count-rate of the flare ($\simeq 100$ cts s$^{-1}$
in the SIS) exceeds both the telemetry limit ($\simeq 40$~cts
s$^{-1}$) and the 1\% pile-up limit throughout the whole point-spread
function ($\simeq 60$ cts s$^{-1}$), thus preventing a reliable
spectroscopic analysis. We therefore only used the GIS data in the
following. Source photons have been extracted, for both GIS-2 and
GIS-3 detectors, from a circular region 18.5~arcmin in diameter (74
pixels, somewhat larger than the suggested extraction radius for GIS
data; given the strength of the source this ensures that as many as
possible source photons are collected) centered on the source
position, while background photons have been extracted from a circular
region identical in size to the source region but symmetrically placed
with respect to the optical axis of the X-ray telescope. For point
sources such strategy allows for the background to be extracted from
the same observation (and thus with the same screening criteria) while
ensuring that the effect of telescope vignetting on cosmic background
photons is properly accounted for.  Given the high intensity of the
source emission during the flare, the background is however
effectively negligible.  The GIS-3 background-subtracted light curve
for EV~Lac for the complete duration of the ASCA observation is shown
in Fig.~\ref{fig:lctotal}.

\begin{figure*}[thbp]
  \begin{center}
    \leavevmode \epsfig{file=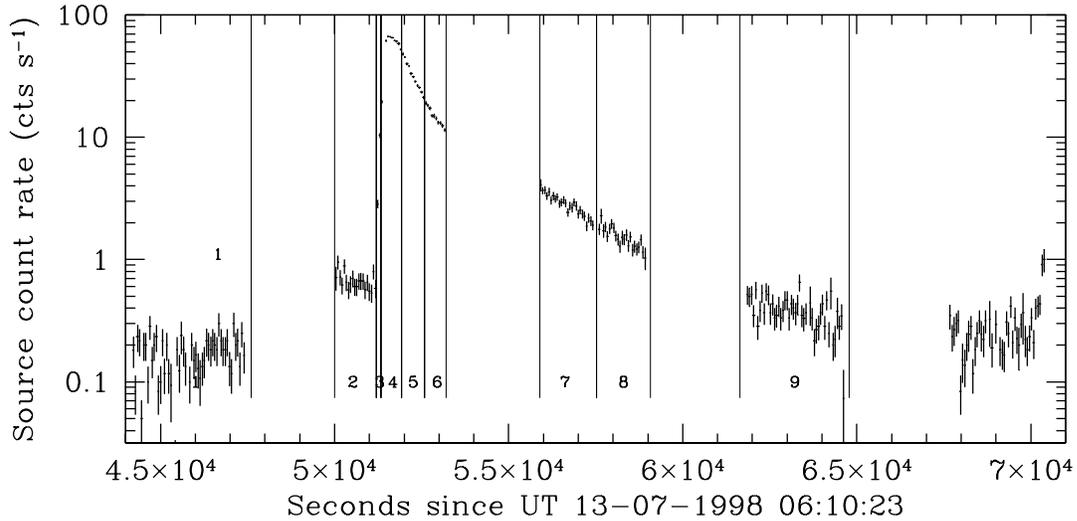, width=15.0cm, bbllx=20pt,
      bblly=400pt, bburx=600pt, bbury=700pt, clip=}
    \caption{The background-subtracted (background count rate $\simeq 0.05$
      cts s$^{-1}$) GIS-3 light curve of the large EV~Lac flare
      discussed in the present paper, binned in 60~s intervals.  The
      extent of the time intervals in which the flare has been divided
      for the determination of the flare's spectral parameters is
      shown. No dead time correction has been applied to the count
      rates.}
    \label{fig:lcflare}
  \end{center}
\end{figure*}

The light curve shows evidence for variability on many time scales,
and at least three individual flares can be identified: the
exceptional event at $\simeq 51$~ks from the beginning of the
observation, and two minor (however still rather sizable) flares at
$\simeq 10$ and $\simeq 75$~ks. The light curves of both minor events
show a clear exponential decay, but their peak is not covered by the
observations. To derive the temporal evolution of the temperature and
emission measure of the large flare we have subdivided it in 9 time
intervals, shown together with the GIS-3 light curve of the event in
Fig.~\ref{fig:lcflare}.  Individual GIS-2 and GIS-3 spectra have been
extracted for each of these intervals and merged using the procedure
described in the \cite*{ascaabc}. The exposure time of each individual
spectrum has been corrected for the dead-time of the GIS (which, at
these high count rates is rather significant, with values up to 1.2 --
note that the both light curves from Figs.~\ref{fig:lctotal}
and~\ref{fig:lcflare} are \emph{not} corrected for detector
dead-time). The quiescent spectrum has been taken from the interval
immediately preceding the flare (interval~1 in Fig.~\ref{fig:lcflare},
covering $\simeq 5$~ks). A two-temperature model has been fit to the
spectrum extracted from interval~1, with the resulting spectral
parameters shown in Table~\ref{tab:quiescent}.

\begin{figure*}[htbp]
  \begin{center}
    \leavevmode \epsfig{file=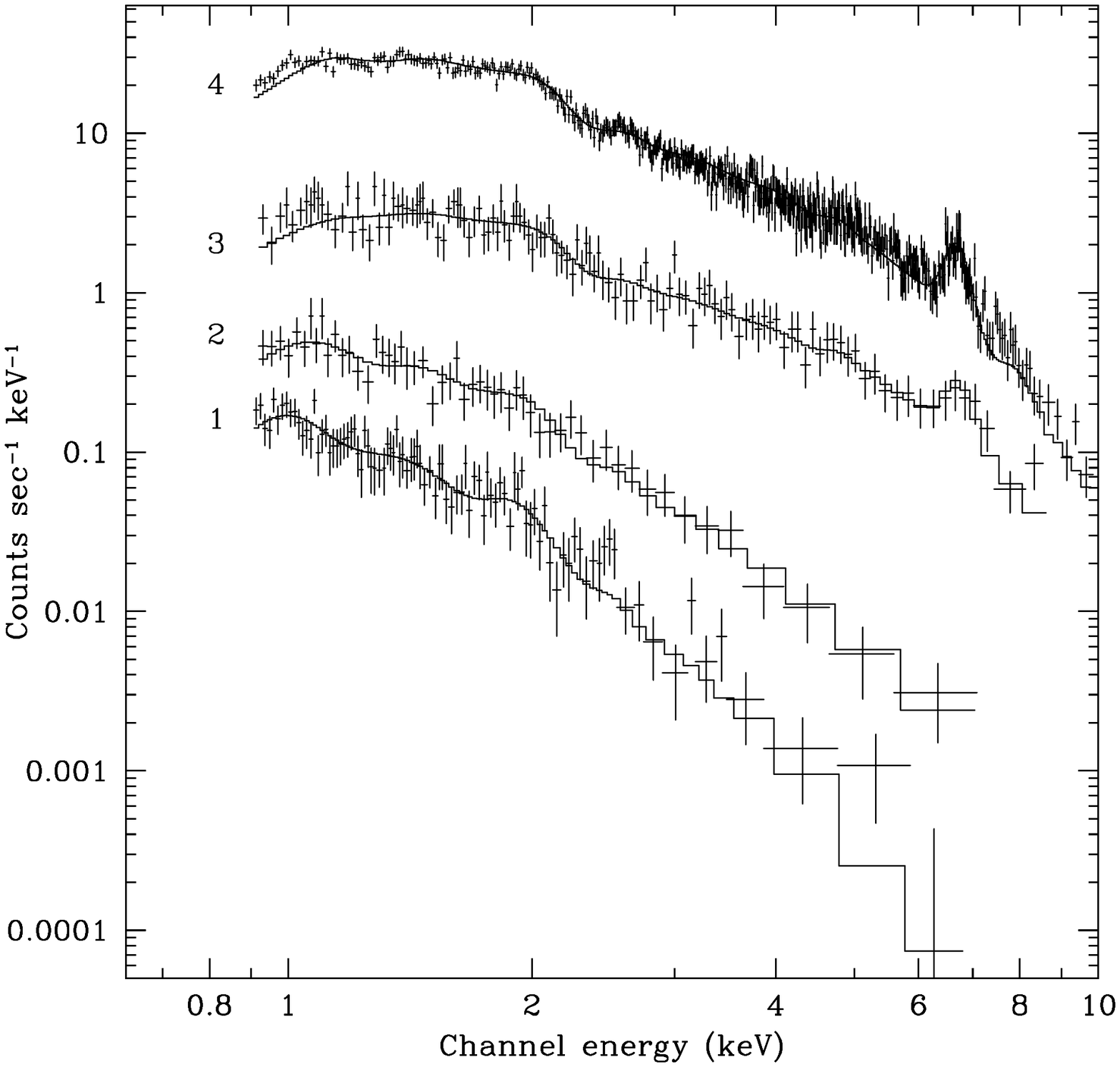, width=8.5cm, bbllx=15pt,
      bblly=150pt, bburx=600pt, bbury=700pt, clip=}
    \leavevmode \epsfig{file=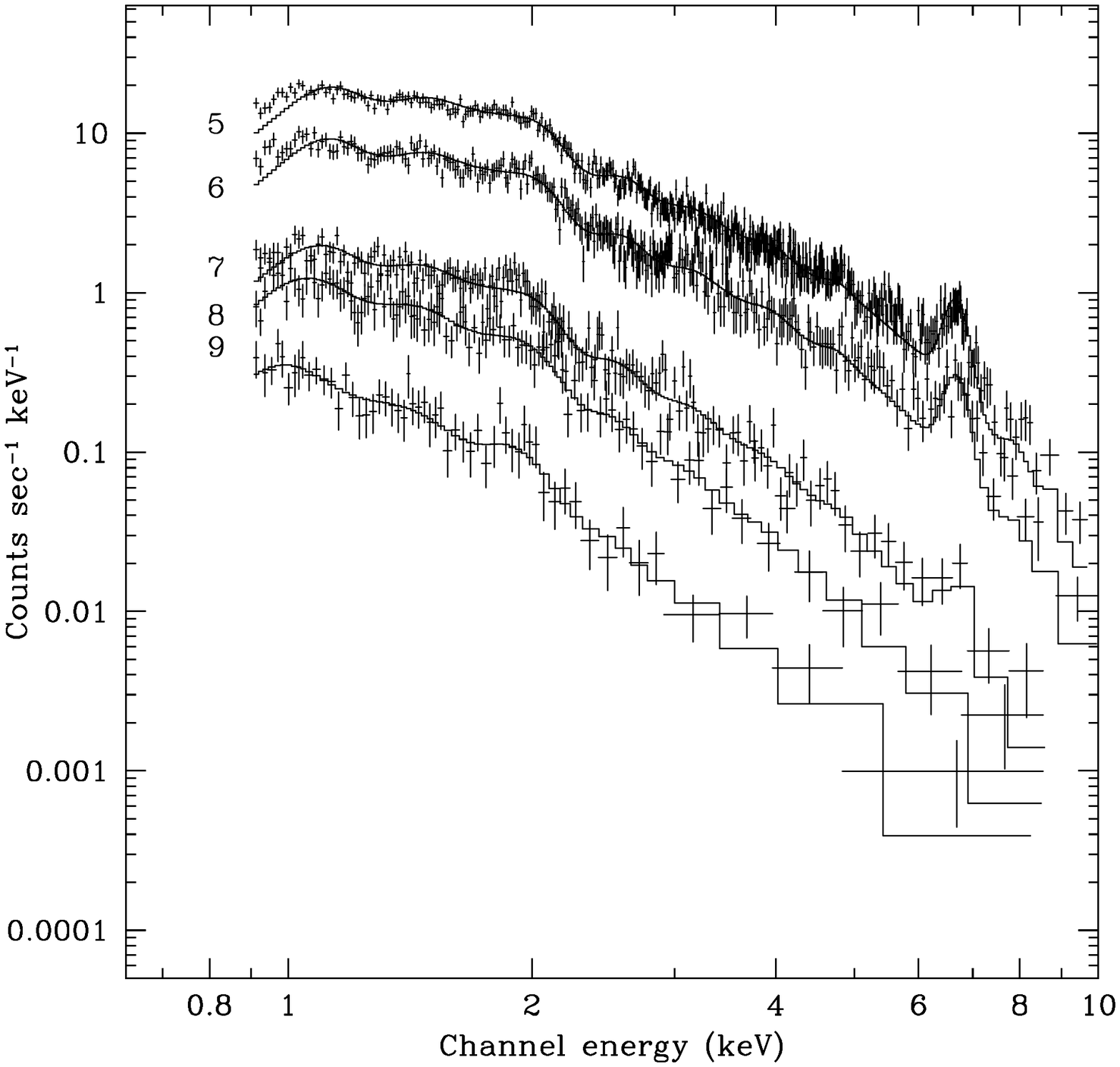, width=8.5cm, bbllx=15pt,
      bblly=150pt, bburx=600pt, bbury=700pt, clip=}
    \caption{The time sequence of merged (GIS-2 $+$ GIS-3) spectra
      for the EV~Lac flare. The left panel shows the sequence of
      spectra during the rise phase of the flare, up to its peak,
      while the right panel shows the sequence of decay spectra. The
      numeric label at the left of each spectrum indicates the time
      interval during which the spectrum was collected, as numbered in
      Fig.~\ref{fig:lcflare}. Each spectrum is overplotted with the
      corresponding best-fit model from which the spectral parameters
      have been derived. The spectra have been rebinned to a minimum
      signal-to-noise ratio per bin of 2.}
    \label{fig:flarespec}
  \end{center}
\end{figure*}

\begin{figure*}[htbp]
  \begin{center}
    \leavevmode \epsfig{file=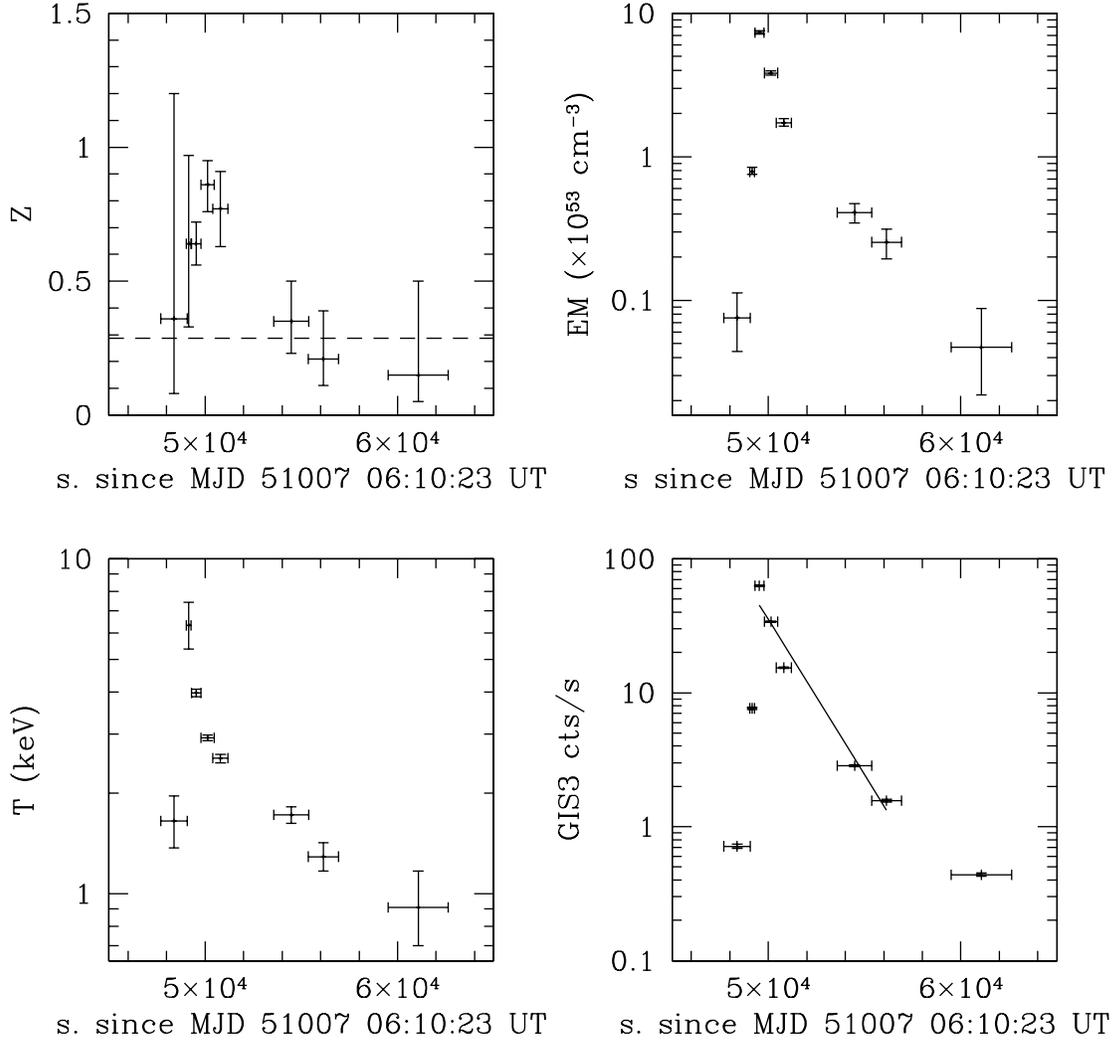, width=15cm, bbllx=25pt,
      bblly=150pt, bburx=600pt, bbury=700pt, clip=}
    \caption{The temporal evolution of the spectral parameters of the
      flare, i.e.\ the plasma metal abundance (top left), emission
      measure (top right) and temperature (bottom left). Also shown
      (bottom right), the light-curve of the flare as seen by the
      GIS-3 detector, integrated in the same temporal intervals as the
      ones used to extract the flare's spectra. The best-fit
      exponential decay to the light curve is also plotted. The dashed
      horizontal line in the abundance plot represents the value of
      the metal abundance determined for the quiescent, pre-flare
      spectrum.}
    \label{fig:para}
  \end{center}
\end{figure*}

\begin{table}[thbp]
    \caption{The spectral parameters derived for the quiescent
      emission of EV~Lac from the analysis of the GIS-2 and GIS-3
      spectra accumulated during the time interval 1, using a
      two-temperature {\sc mekal} spectral model. The fit has 72
      degrees of freedom, and the corresponding probability level is
      99.8\%. The quiescent X-ray luminosity corresponding to the
      above spectral parameters is $L_{\rm X} = 3 \times
      10^{28}$~erg~s$^{-1}$.}
  \begin{center}
    \leavevmode
    \begin{tabular}{cc|cc|c|c}
 $T_1$ & $T_2$ &  $EM_1$ & $EM_2$  &  $Z$ & $\chi^2$ \\
\multicolumn{2}{c|}{keV}  & \multicolumn{2}{c|}{$10^{51}$ cm$^{-3}$} &
$Z_\odot$  & \\\hline
 & & & & & \\[0.5pt]
$0.78\pm 0.15$ & $1.92 \pm 1.2$ & 2.35 & 0.76 & 0.29 & 0.97 \\
    \end{tabular}
    \label{tab:quiescent}
  \end{center}
\end{table}

The sequence of GIS flare spectra is shown in
Fig.~\ref{fig:flarespec}. The left panel shows the spectra collected
during the rising phase of the flare, up to the peak in the
light-curve (interval~4), while the decay-phase spectra are plotted in
the right panel.  Individual flare spectra from time intervals 2 to 9
have been fit with a single-temperature {\sc mekal} model; given the
lack of soft response in the GIS no absorbing column density was
included. The quiescent emission was modeled by adding a
frozen-parameter two-temperature {\sc mekal} model to the fit (with
the parameters as in Table~\ref{tab:quiescent}); the results of the
spectral fits are shown in Table~\ref{tab:flaring}. During intervals 5
and 6 the single-temperature fit does not yield a satisfactory reduced
$\chi^2$ (see Table~\ref{tab:flaring}), due to the large residuals
present in the region around 1~keV, with the observed spectra showing
some additional emissivity with respect to the models.  Similar
effects are also seen in the in the ASCA SIS Capella spectrum
(\cite{bri98}) and in the flaring spectra of Algol as seen by SAX
(\cite{fs99}), suggesting that current plasma emission codes (as the
{\sc mekal} one used here) under-predict the observed spectrum
around 1~keV, likely due to a large number of high quantum-number
($n>5$) Fe~L lines from Fe\,{\sc xvii}, Fe\,{\sc xviii} and Fe\,{\sc
  xix} (\cite{bri98}), and thus the formally unacceptable $\chi^2$
resulting from the fit does not necessarily imply that the
one-temperature model is not correctly describing the observed
spectrum. Indeed, we have verified that adding further temperature
components does not significantly improve the fit for intervals 5 and
6. The time evolution of the flare's spectral parameters (temperature,
emission measure, plasma metal abundance) is shown in
Fig.~\ref{fig:para}, together with the flare's GIS-3 light-curve
binned in the same time intervals.

\begin{table*}[htbp]
  \caption{The spectral parameters $T$, $Z$ and $EM$ derived for the
    individual phases of the EV~Lac flare from the analysis of the GIS
    spectra accumulated during the time intervals 2 to 9. The spectra
    have been analyzed with a single-temperature {\sc mekal} model
    (plus a frozen-parameter two-temperature model to account for the
    quiescent emission).  The bounds of the confidence intervals at
    the 90\% levels ($\Delta \chi^2 = 3.5$) are also reported for each
    parameter. The time interval $i$ to which each set of parameters
    applies is shown in Fig.~\ref{fig:lcflare}.}
  \begin{center}
    \leavevmode
    \begin{tabular}{r|rrr|rrr|lll|rrr}
      $i$ & $T$ & $T_{+90\%}$ & $T_{-90\%}$  &
      $Z$ & $Z_{+90\%}$ & $Z_{-90\%}$ &  $EM$ & $EM_{+90\%}$ & $EM_{-90\%}$
      & $\chi^2$ & DoF & Prob.\\ 
      & \multicolumn{3}{c|}{keV}  & \multicolumn{3}{c|}{$Z_\odot$} &
      \multicolumn{3}{c|}{$10^{53}$ cm$^{-3}$} & 
      & & \\\hline
      2 & 1.65 & 1.37 & 1.96 & 0.36 & 0.08 & 1.20 & 0.075 & 0.044 & 0.11 &
      30.1 & 57 & 0.995 \\ 
      3 & 6.33 & 5.37 & 7.42 & 0.64 & 0.33 & 0.97 & 0.79 & 0.75 & 0.85 &
      88.1 & 128 & 0.995 \\
      4 & 3.98 & 3.88 & 4.08 & 0.64 & 0.56 & 0.72 & 7.35 & 7.19 & 7.54 &
      481.0 & 468 & 0.294 \\
      5 & 2.92 & 2.86 & 2.98 & 0.86 & 0.76 & 0.95 & 3.83 & 3.71 & 3.96 &
      561.2 & 417 & $<0.01$ \\
      6 & 2.54 & 2.46 & 2.61 & 0.77 & 0.63 & 0.91 & 1.73 & 1.63 & 1.85 &
      391.1 & 297 &  $<0.01$ \\
      7 & 1.72 & 1.62 & 1.82 & 0.35 & 0.23 & 0.50 & 0.41 & 0.35 & 0.47 &
      190.1 & 183 &  0.27 \\
      8 & 1.29 & 1.17 & 1.42 & 0.21 & 0.11 & 0.39 & 0.25 & 0.20 & 0.31 &
      99.0 & 117 &  0.84 \\
      9 & 0.91 & 0.70 & 1.17 & 0.15 & 0.05 & 0.50 & 0.047 & 0.022 & 0.088 &
      50.8 & 67 & 0.88  \\
    \end{tabular}
    \label{tab:flaring}
  \end{center}
\end{table*}

\section{Flare analysis}
\label{sec:flare}

We have analyzed the present flare using two different frameworks
(quasi-static cooling and hydrodynamic decay, sustained heating),
which both make the assumption that the flaring plasma is confined in
a closed loop structure, whose geometry is not significantly evolving
during the event. Although direct support to this assumption is of
course missing, the relatively short duration of the event allows an
analogy with solar compact flares. In any case the second method
provides reliable scale sizes of the flaring structure even in the
presence of some readjustment of the magnetic field, the crucial
assumption been plasma confinement.

\subsection{The quasi-static cooling framework}

Many stellar X-ray flares observed to date have been studied using the
quasi-static formalism first discussed in detail by \cite*{om89}.  It
is thus of interest to analyze the present event with the quasi-static
approach, to allow a homogeneous comparison with literature data, even
if, as discussed by \cite*{fs99}, this method can significantly
over-estimate the size of the flaring loops (see also \cite{rbp+97}).

According to \cite*{om89}, for the decay to be quasi-static (i.e.\ to
happen through a sequence of states each of which can be described by
the scaling laws applicable to stationary coronal loops) the ratio
between the characteristics times for radiative and conductive cooling
must be constant during the flare decay (although its absolute value
needs not be known). This ratio is parameterized by the quantity

\begin{center}
  \begin{equation}
    \label{eq:mu}
    \mu = {\tau_{\rm r} \over \tau_{\rm c}} = C \times {T^{13/4}
      \over {EM}} , 
  \end{equation}
\end{center}

\begin{figure}[htbp]
  \begin{center}
    \leavevmode \epsfig{file=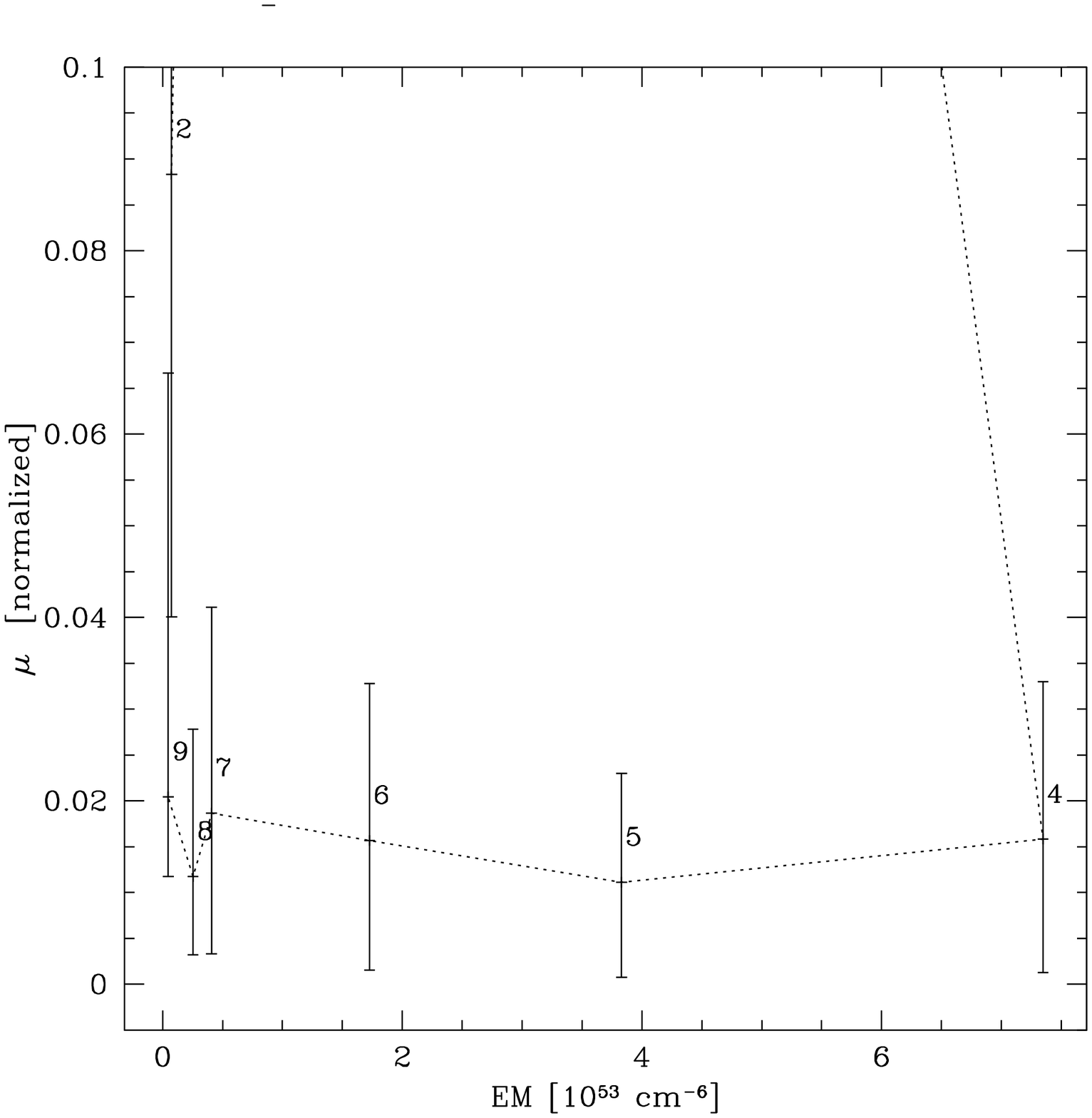, width=8.5cm, bbllx=15pt,
      bblly=150pt, bburx=600pt, bbury=700pt, clip=}
    \caption{The temporal evolution of the $\mu$ parameter (the ratio
      between the radiative and conductive cooling time for the loop)
      during the flare. The dotted line joins the points corresponding
      to the different time intervals reported in
      Fig.~\ref{fig:lcflare}.}
    \label{fig:mu}
  \end{center}
\end{figure}

The normalization $C$ depends on the details of the loop's geometry,
and is not relevant here. The evolution of $\mu$ is plotted in
Fig.~\ref{fig:mu}; within the error bars $\mu$ is constant during the
whole decay, so that the conditions for the applicability of the
quasi-static decay framework are in this case met.

The scaling laws discussed by \cite*{sut+92} --linking the loop
semi-length $L$ and the plasma density $n$ with the peak flare
temperature $T$ and the effective decay time $\tau$ -- yield results
very similar to the ones obtained through a full fit to the equations
of \cite*{om89}, so that we will limit ourselves to their application.
Specifically, $L \propto \tau T^{7/8}$ and $n \propto \tau^{-1}
T^{6/8}$, valid for temperature regimes in which the plasma emissivity
scales as $\Psi_0 T^{-\gamma}$ with $\gamma \simeq -0.25$.  In
practice this is valid for $T \ga 20$\,MK, i.e.\ for most of the decay
of the flare discussed here.  Scaling the values determined for the
EV~Lac flare from the parameters derived by \cite*{om89} for the
EXOSAT flare observed on Algol, the derived loop semi-length is $L
\simeq 5 \times 10^{10}$~cm ($\simeq 2~R_*$), and the plasma density
$n \simeq 6 \times 10^{11}$ cm$^{-3}$.

\subsection{The hydrodynamic decay, sustained heating framework}

A different approach to the study of a flare's decay phase, with the
same aim of determining the physical parameters of the flaring region,
has been developed by \cite*{rbp+97}. It has been recognized from the
modeling of solar X-ray flares that the slope of the locus of the
flare decay in the $\log n$--$\log T$ plane is a powerful diagnostic
of the presence of additional heating during the decay itself
(\cite{sss+93}); by making use of extensive hydrodynamic modeling of
decaying flaring loops, with different heating time scales,
\cite*{rbp+97} have derived an empirical relationship between the
light curve decay time (in units of $\tau_{\rm th}$, the loop
thermodynamic decay time, \cite{srj+91}) and the slope $\zeta$ in the
$\log n$--$\log T$ diagram. This allows to derive the length of the
flaring loop length as a function of observable quantities, i.e.\ the
decay time of the light curve, the flare maximum temperature and the
slope of the decay in the $\log n$--$\log T$ diagram (the square root
of the emission measure of the flaring plasma is actually used as a
proxy to the density).  Since the characteristics of the observed
decay depend on the specific instrument response, the parameters of
the actual formulas used have to be calibrated for the telescope used
to observe the flare.

The method reported in \cite*{rbp+97} was tested on a sample of solar
flares observed with Yohkoh-SXT, for which both images (from which the
length of the flaring loop was measured) and spectral parameters (from
the temperature and emission measure diagnostic derived from Yohkoh
filter ratios) were available, and has been shown to provide reliable
results for most of the studied events.  A first application of the
method to stellar flares observed with ROSAT PSPC is described by
\cite*{rm98}.

For the present study the method has been recalibrated for stellar
flares observed with ASCA GIS. The thermodynamic decay time $\tau_{\rm
  th}$ of a closed coronal loop with semi-length $L$, and maximum
temperature $T_{\rm max}$ is given by \cite*{srj+91} as

\begin{equation}
  \label{eq:taulc}
  \tau_{\rm th} = {\alpha L \over \sqrt{T_{\rm max}}}
\end{equation}
where $\alpha = 3.7 \times 10^{-4} \rm ~ cm^{-1}~s^{-1}~K^{1/2}$. By
means of a grid of hydrostatic loop models (see \cite{rm98}) we have
determined an empirical relationship linking the loop maximum
temperature $T_{\rm max}$, typically found at the loop apex (e.g.\ 
\cite{rtv78}) to the maximum temperature $T_{\rm obs}$ determined from
the GIS spectrum:

\begin{equation}
  \label{eq:tobs}
  T_{\rm max} = 0.085 \times T_{\rm obs}^{1.176}
\end{equation}

Following the same procedure as in \cite*{rbp+97} and \cite*{rm98}
(using extensive hydrodynamic modeling of decaying flaring loops) we
have determined the ratio between $\tau_{\rm LC}$ (the observed
$e$-folding time of the flare's light curve determined by fitting the
light curve from the peak of the flare down to the 10\% of peak level)
and $\tau_{\rm th}$ as a function of the slope $\zeta$ in the $\log
\sqrt{EM}$--$\log T$ diagram to be, for the ASCA GIS

\begin{equation}
  \label{eq:zeta}
  {\tau_{\rm LC} \over \tau_{\rm th}} = F(\zeta) =
  {c_a}e^{-\zeta/\zeta_a} + q_a  
\end{equation}
where $c_a = 10.9$, $\zeta_a = 0.56$ and $q_a = 0.6$.  The formula for
the loop semi-length $L$ is therefore:

\begin{equation}
  \label{eq:lzeta}
  L=\frac{\tau_{\rm LC} \sqrt{T_{\rm max}}}{\alpha F(\zeta)}  ~~~~~~~~ 0.38 <
  \zeta \leq 1.7 
\end{equation}
where the second part of the relationship gives the range of $\zeta$
values allowed according to the modeling. The uncertainty on $L$ is
the sum of the propagation of the errors on the observed parameters
$\tau_{\rm LC}$ and $\zeta$ with the standard deviation of the
difference between the true and the derived loop lengths. The latter
amounts to $\simeq 15$\%. Equation \ref{eq:lzeta} has been tuned on
exponentially decaying light curves; however it has been shown to
provide reliable results also on solar flares with more complex decay
trends, e.g. a double exponential decay (as for the flare studied
here), provided that the whole decay is fitted with a single
exponential (F.~Reale, private communication).

\begin{figure}[htbp]
  \begin{center}
   \leavevmode \epsfig{file=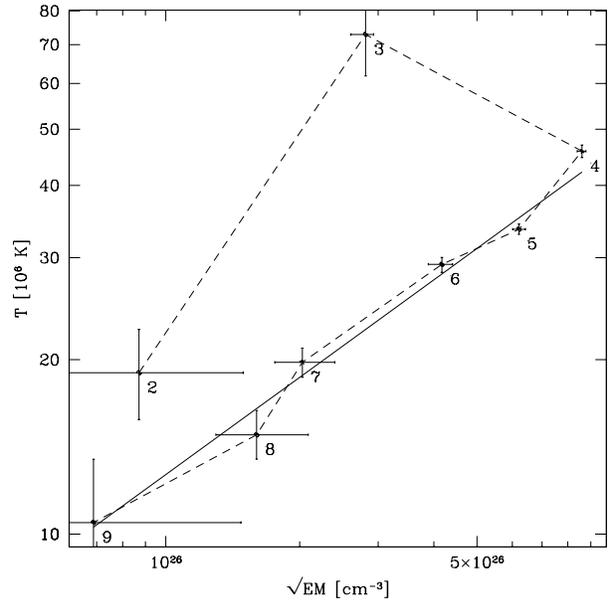, width=8.5cm, bbllx=25pt,
     bblly=150pt, bburx=600pt, bbury=700pt, clip=}
    \caption{The evolution of the flare's decay in the $\log
      \sqrt{EM}$--$\log T$ plane. The points plotted represent the
      flare's evolution from time interval 2 to time interval 9
      inclusive. The dotted line joins the points corresponding to
      successive intervals. The decay phase begins with the third
      interval, and it closely follows a straight line. The continuous
      line is a least-square fit to the decay phase, with a best-fit
      slope $\zeta = 0.56 \pm 0.04$, while the numeric labels indicate
      the time interval to which each point refers.}
    \label{fig:zeta}
  \end{center}
\end{figure}

The evolution of the EV~Lac flare in the $\log \sqrt{EM}$--$\log T$
plane is shown in Fig.~\ref{fig:zeta}, together with a least-square
fit to the decay phase. The resulting best-fit slope for the decaying
phase computed for time intervals from 4 to 9 inclusive is $\zeta =
0.56 \pm 0.04$. Application of Eq.~\ref{eq:zeta} yields a ratio
between the observed cooling time scale $\tau_{\rm LC}$ and the
thermodynamic cooling time scale for the flaring loop $\tau_{\rm th}$
of $F(\zeta) = 4.6$. Such a large value implies that the observed
decay is driven by the time-evolution of the heating process and not
by the free cooling of the loop.  Also, the actual loop length will be
significantly smaller than it would be estimated assuming that the
spectral parameters reflect free cooling of the decaying loop. The
actual value of $\tau_{\rm LC}$ has been determined by fitting the GIS
light curve, binned in the same time intervals used for extracting the
individual flare spectra (as plotted in Fig.~\ref{fig:lcflare}),
considering the intervals from 4 to 9 inclusive.  In this case
$\tau_{\rm LC} = 1.80 \pm 0.15$~ks, and therefore $\tau_{\rm th}
\simeq 400$~s. The intrinsic flare peak temperature is, applying
Eq.~\ref{eq:tobs} to the observed maximum temperature, $T_{\rm max}
\simeq 150$~MK. From Eq.~\ref{eq:lzeta} the loop semi-length is
$L=(1.3 \pm 0.3) \times 10^{10}$~cm, i.e.\ $L \simeq 0.5~R_*$. This
loop length is much smaller than the pressure scale
height\footnote{defined as $H = 2kT/\mu g$, where T is the plasma
  temperature in the loop, $\mu$ is the molecular weight and $g$ is
  the surface gravity of the star. In this case $H \simeq 8 \times
  10^{11}$~cm} of the flaring plasma on EV Lac and also significantly
smaller (by a factor of 4) than the one derived through the
quasi-static formalism.

A simple consistency check can be obtained by comparing the pressure
obtained by assuming that the flaring loop is not, at maximum, far
from a steady-state condition (thus applying the scaling laws of
\cite{rtv78}) with the pressure implied by the derived values of $L$
and $T$ for a plausible geometry. In practice the geometry is
parameterized by the ratio $\beta$ between the radius of the loop and
its lenght. The pressure is then

\begin{equation}
  \label{eq:dens}
  n = \sqrt{{ {EM} } \over {2 \pi L^3 \beta^2}}
\end{equation}

If we assume $\beta \simeq 0.1$--$0.3$ (a typical range for solar
coronal loops) the loop volume is $\simeq 1.4$--$13 \times
10^{29}$~cm$^3$, and the resulting plasma density at the peak of the
flare is $n \simeq 2$--$0.2 \times 10^{12}$~cm$^{-3}$. The
corresponding pressure is $p_{\rm max} \simeq 8$--$0.9 \times
10^4$~dyne~cm$^{-2}$. Using the scaling laws of \cite*{rtv78}
applicable to steady-state loops,

\begin{equation}
\label{eq:rtv}
T_{\rm max} = 1.4 \times 10^3 (p_0 L)^{1/3}
\end{equation}
where $p_0$ is the pressure at the base of the loop, one obtains $p_0
\simeq 10^5$~dyne~cm$^{-2}$, slightly larger than $p_{\rm max}$ for
$\beta=0.1$. This implies that the plasma evaporated from the
chromosphere has not ($\beta = 0.3$) or nearly ($\beta = 0.1$) filled
the flaring loop up to the hydrostatic equilibrium conditions at flare
maximum.

\subsection{Energetics}
\label{sec:energy}

\begin{figure}[thbp]
  \begin{center}
    \leavevmode \epsfig{file=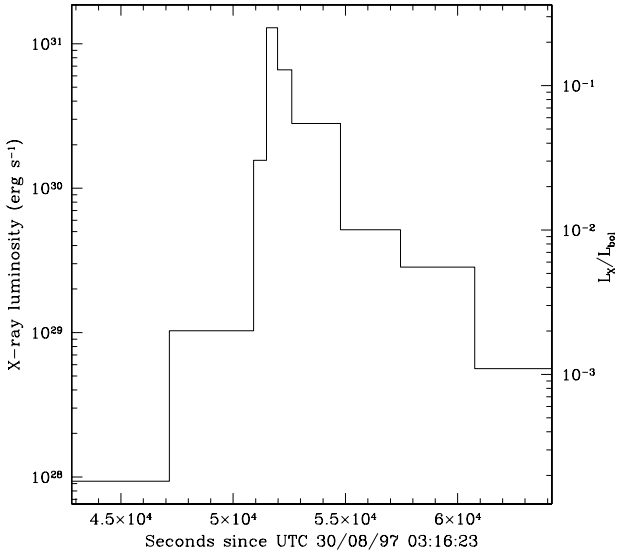, width=8.5cm, bbllx=15pt,
      bblly=150pt, bburx=600pt, bbury=700pt, clip=}
    \caption{The temporal evolution of the flare's X-ray luminosity
      calculated in the 0.1 to 10.~keV band starting from the
      quiescent state. The left axis gives the luminosity in
      erg~s$^{-1}$, while the right axis shows the same value in units
      of the star's bolometric luminosity ($5.25 \times 10^{31}$
      erg~s$^{-1}$).}
    \label{fig:lum}
  \end{center}
\end{figure}

We have computed, for each of the 8 time intervals in which the flare
has been subdivided, the X-ray luminosity in the 0.1--10.~keV band.
For this purpose the spectrum has been extrapolated outside of the
formal spectral range covered by the GIS detectors; this is at most
likely to underestimate the true luminosity in the extended band, as
it may miss any softer component present in the spectrum and invisible
to the GIS. The time-evolution of the flare X-ray luminosity is shown
in Fig.~\ref{fig:lum}, in which the data are plotted both in absolute
units and in units of the star's bolometric luminosity.

During interval 4 (at the peak of the light curve) the X-ray
luminosity of the flare is about one quarter of the photospheric
(bolometric) luminosity of the star. Soft X-ray radiation is only one
of the energy loss terms for the flaring plasma, with kinetic energy,
conduction to the chromosphere and white light, UV and XUV flaring
emission also significantly contributing to the energy budget. In the
solar case, detailed analyses of flares of different types
(\cite{wjd+86}) show that at the peak of the event soft X-ray
radiation only accounts for 10--20\% of the total energy budget;
similarly, \cite*{hfd+93} analyzed a large optical flare on the dMe
star AD~Leo, concluding that the kinetic energy of plasma's motions
during the event is likely to be at least as large as the radiated
energy during the flare.

A detailed assessment of the energy balance of the present flare is
not possible, given the lack of multi-wavelength coverage and of
velocity information which could help assessing the plasma kinetic
energy. The total energy radiated in the X-rays is $E \simeq 1.5
\times 10^{34}$~erg (obtained with simple trapezoidal integration of
the instantaneous X-ray luminosity), over $\simeq 10$~ks, equivalent
to $\simeq 300$~s of the star's bolometric energy output.  From the
scaling laws of \cite*{rtv78} we can estimate the heating rate per
unit volume at the peak of the flare, assumed uniform along the loop,
as

\begin{equation}
  \label{eq:rtvh}
  \frac{d H}{d V d t} \simeq 10^5 ~ p^{7/6} ~ L^{-5/6} \simeq 240 ~
  \rm erg ~ cm^{-3} ~ s^{-1}
\end{equation}

The total heating rate at the flare maximum is therefore

\begin{equation}
  \frac{d H}{dt} \simeq \frac{d H}{d V d t} \times V \simeq 3.3 \times
  10^{31} ~ \rm erg ~ s^{-1}
\end{equation}
a factor of $\simeq 3$ higher than the flare maximum X-ray luminosity
(see Fig.~\ref{fig:lum}), compatible with X-ray radiation being only
one of the energy loss terms during the flare. If we assume that the
heating is constant for the initial rising phase, which lasts for
$t_{\rm r} \simeq 300$ s, and then decays exponentially, with an
$e$-folding time $\tau_{\rm H} \simeq 4.6 \tau_{\rm th} \simeq 1800$ s
(i.e.\ similar to $\tau_{\rm LC}$), the total energy released during
the flare is

\begin{equation}
  H_{\rm tot} \simeq \frac{d H}{dt} \times (t_r + \tau) \simeq 7
  \times 10^{34} ~ \rm erg
\end{equation}
approximately five times the energy radiated in X-rays. Energy losses
by thermal conduction are indeed expected to be large at such high
temperatures.

\subsection{Metal abundance}

The metal abundance of the flaring plasma shows a significant
evolution, rising, at the peak of the flare, to a value $\simeq 2$
times higher than the abundance measured for the quiescent emission,
and decaying back to the quiescent value during the terminal phase of
the flare. Evolution of the metal abundance has been observed also in
other flares -- for example it is evident in the Algol flare observed
with SAX, \cite*{fs99} -- and the behavior observed here appears to
follow the same general pattern of abundance enhancement going in
parallel with the flare's light curve. The simplest explanation
suggested for this has been to assume that a fractionation mechanism
is at work in the quiescent corona, and that the evaporation of
pristine photospheric material during the flare temporarily sets the
coronal chemical equilibrium off balance.  Unfortunately, few reliable
abundance determinations are available for M dwarfs; in particular, no
state-of-the-art photospheric abundance analysis of EV~Lac is known to
us, and thus it is not possible to assess whether the quiescent
coronal abundance is indeed depleted with respect to the photospheric
value.

\section{Discussion}
\label{sec:disc}

The most remarkable characteristic of the EV~Lac flare discussed in
the present paper is certainly its very large X-ray luminosity: at
flare peak, for a few minutes, the flare nearly outshines the star's
photosphere. A detailed analysis of the flare's decay, however, shows
that this is an interesting event also on other accounts.  Typical
loop lengths derived for strong flares on active stars, mostly using
the quasi-static cooling mechanism, are large, comparable or often
greater than the stellar radius. The picture which has emerged from
most of the literature is thus one of long, tenuous plasma structures,
with the attendant challenges of sufficiently strong magnetic fields
far away from the stellar surface.  In the present case, instead, the
loop semi-length derived from the analysis of the flare decay using
the method of \cite*{rbp+97} is relatively compact, at about 0.5
stellar radii (implying a maximum height above the stellar surface of
$\simeq 0.3$ stellar radii). The length derived using the quasi-static
formalism is about 4 times larger, and would thus again lead to the
``classic'' picture of long, tenuous loops.

Although certainly not small, flaring loops of $L \simeq 0.5~R_*$ are
by no means exceptional, even by the relatively modest solar
standards.  The analysis of the large flare on Algol seen by the SAX
observatory (\cite{sf99}; \cite{fs99}) shows that the picture of large
tenuous loops implied by the results of the quasi-static analysis can
be quite misleading, and that, at least in that case, the geometrical
size of the flaring plasma constrained by the light-curve eclipse is
significantly smaller than the loop heights derived with the
quasi-static method. An analysis based on the method of \cite*{rbp+97}
yields, also in that case, loop dimensions which are significantly
smaller than the ones implied by the quasi-static analysis. The much
more compact loop size derived through the \cite*{rbp+97} method is
linked with the presence of significant plasma heating during the
flare decay, as the implied small, dense loop has a very short
thermodynamic decay time ($\tau_{\rm th} \simeq 400$~s). No large
diffuse plasma structures at several stellar radii from the surface
are needed to model the flaring region, and a more appropriate picture
appears to be one of a rather compact, high-pressure plasma structure,
whose decay is completely dominated by the time-evolution of the
heating mechanism.

The present large EV~Lac flare shows several characteristics in common
with other large, well observed stellar flares discussed in the
literature. The light curve has a ``double exponential'' decay, with
the initial time scale being more rapid, and a slower decay setting in
afterwards. A very similar decay pattern is observed in the large
Algol flare seen by SAX as well as on many large solar flares (see
\cite{fsd+95} for an example). The best-fit metal abundance for the
flaring plasma also shows what by now appears to be a characteristic
behavior, i.e.\ it increases in the early phases of the flare, peaking
more or less with the peak of the light-curve, and then it decreases
again to the pre-flare levels. In the case of the SAX Algol flare the
long duration of the event and the high intensity of the flare make it
possible to show that the metallicity decays to its pre-flare value on
faster time scales than either the plasma temperature or the emission
measure.  The coarser time coverage of the present flare and the
shorter duration of the event do not allow such detailed assessment.

The heating mechanism of the solar (and stellar) corona remains in
many respects an unsolved puzzle, and even more the mechanism
responsible for large flares.  However, it is by now rather clear that
most sizable flares cannot be explained with a simple picture of a
(mostly) impulsive heating event followed by decay dominated by the
free thermodynamic cooling of the plasma structure. On the contrary,
the evidence from the recent flow of well studied flare data
(including the one in the present paper) is that the decay of large
flares is dominated by the time evolution of the heating mechanism.
Thus, the double exponential decay observed here as well as in other
large solar and stellar flare is likely to be a characteristic of the
heating mechanism rather than one of the flare decay.

The interpretation of the increase in best-fit plasma metallicity
during the flare's peak is still unclear. If the presence of a
fractionation mechanism is accepted, which causes differences in the
metal abundances in the photosphere and in the corona, the abundance
increase during the flare can plausibly be explained as due to the
evaporation of photospheric plasma during the early phases of the
flare, on time scales faster than the ones on which the fractionation
mechanism operates. Since the coronal plasma shortly after the
impulsive heating is almost entirely made of material evaporated from
the chromosphere, if this scenario were correct the observations
presented here would imply that the chromospheric metallicity should
be about three times the coronal one in quiescent conditions.

Recent calculations (G.~Peres, private communication) show that the
plasma in a flaring loop such as the one responsible for the EV~Lac
flare discussed here is not optically thin for the strongest lines.
This is in particular true for the Fe\,{\sc xxv} complex at $\simeq
6.7$~keV, which drives the determination of the metallicity of the
flaring plasma. However, optical thickness effects would in general
depress the strong line, leading to a lower metallicity estimate, and
cannot therefore explain the metallicity increase observed during the
flare. Another possible bias to the best-fit metallicity can derive
from the thermal structure of the flaring plasma, which is not
isothermal, even if it's being fit with an isothermal model. To assess
the magnitude of this effect we analyzed the synthetic spectra
produced with an ad hoc hydrodynamic simulation of a flaring loop,
showing that the effect is small ($\le 30$\%) in comparison with the
observed magnitude of the change (a factor of $\simeq 3$).

If the heating mechanism responsible for the present flare is
essentially due to some form of dissipation of magnetic energy, an
obvious question to ask is what field strength would be required to
accumulate the emitted energy, and to keep the plasma confined in a
\emph{stable} magnetic loop configuration.  A related question is
whether the present flare could be interpreted with a reasonable
scaling of the conditions usually observed in the solar corona, or
whether a different configuration and/or mechanism for energy release
need to be invoked.  Our flare analysis allows to make some relevant
estimates, under the assumptions that the energy release is indeed of
magnetic origin and it occurs entirely within a single coronal loop
structure, with the characteristics inferred from the analysis of the
flare decay.  An estimate of the minimum magnetic field $B$ necessary
to produce the event can then be obtained from the relation:

\begin{center}
  \begin{equation}
    \label{eq:b}
    E_{\rm tot} = {{(B^2 - B_{\rm 0}^2) \times V_{\rm loop}} \over 8 \pi}
  \end{equation}
\end{center}
where $E_{\rm tot} \simeq 7 \times 10^{34}$~erg is the total energy
released (Sect.~\ref{sec:energy}), $B_{\rm 0}$ is the magnetic field
surviving the flare and $V_{\rm loop}$ is the total volume of the
flaring plasma. As an estimate of $B_{\rm 0}$ we take the magnetic
field necessary to maintain the plasma confined in a rigid loop
structure along the whole flare, thus implicitly assuming that the
loop geometry does not change during the flare.  From a plasma density
$n \simeq 2 \times 10^{12}$~cm$^{-3}$, and a temperature $T \simeq
100$~MK the estimated maximum plasma pressure is $6 \times
10^4$~dyn~cm$^2$; in order to support such a pressure, a field of
$\simeq 1.2$~kG is required.  Hence, the total minimum magnetic field
required to explain the flare is, from Eq.~\ref{eq:b}, $B \simeq
3.7$~kG, a value compatible with the average magnetic field of 3.8~kG,
with a surface filling factor of about 60\% (and evidence for field
components of up to $\simeq 9$~kG), measured on EV~Lac by \cite*{jv96}
at photospheric level.

Although we have used the loop volume in the derivation of $B$, this
is not to say that the field fills up the whole volume.  Rather, our
estimates can be interpreted in the framework of the flare energy
stored in a magnetic field configuration (e.g.\ a large group of
spots) with a field strength of several kG, covering a volume
comparable to the one of the flaring loop.  What rests as a matter of
speculation is how often such a large energy release may occur, or in
other words, what are the conditions required to accumulate such large
amounts of magnetic energy, especially when the photosphere is so
permeated of magnetic fields as shown by \cite*{jv96}.

In any case, the above scenario suggests the presence of large-scale,
organized magnetic fields. This is somewhat in contrast with the
hypothesis that EV Lac is a fully convective star, whose activity is
powered by a turbulent dynamo, which would be expected to produce
small-scale magnetic fields. Most dynamo theories suggest
(\cite{dyr93}) that less magnetic flux should be generated by a
turbulent dynamo (as compared to the case of the solar-type ``shell''
dynamo) because there is no stable overshoot layer where the fields
can be stored and amplified, and only small-scale magnetic regions
should emerge uniformly to the surface, because the crucial ingredient
is small-scale turbulent flow field, rather than large-scale
rotational shear.

On the other hand the presence of a magnetic field may substantially
modify the stellar interior structure. Magnetic fields -- even smaller
than dictated by equipartition arguments -- alter the convective
instability conditions (\cite{vzm+98}), and thus likely modify the
structure of the convective envelope. At the same time convection has
the tendency to pump magnetic fields downward (``turbulent pumping'',
\cite{bmt92}; \cite{tbc+98}), so that -- in a fully-convective star --
fields may accumulate near the center. Hence, magnetic fields are
likely to be an important (and thus far essentially unaccounted) term
in determining the actual stellar structure, and any realistic
calculation at the low-mass end should consider the dynamo-generated
magnetic fields as an essential part.  Indeed, the possibility that a
strong magnetic field can lead to the formation of a radiative core
has been discussed by \cite*{csh81}, and this may be the seed for a
resurrection of a ``shell'' dynamo mechanism.

\begin{acknowledgements}
 
  We would like to thank J.~Schmitt for the helpful discussion
  relative to the choice of the best target for this observing
  program, and G.~Peres and S.~Orlando for the illuminating
  discussions on many details of flare evolution. FR, GM, SS and AM
  acknowledge the partial support of ASI and MURST.

\end{acknowledgements}


\appendix

\section{Physical characteristics of EV~Lac}
\label{sec:targ}

EV~Lac (Gl 873) is a dMe dwarf, classified as M3.5 (\cite{rhj95}), at
a distance (from the Hipparcos-measured parallax) $d = 5.05$~pc. It is
considered a single star, with no evidence of companions, and is a
slow rotator, with a photometrically determined rotation period of
4.378~d (\cite{pos92}). The projected equatorial velocity has been
determined from Doppler broadening of the spectral lines at $v \sin i
= 4.5 \pm 0.5$~km~s$^{-1}$ (\cite{jv96}) and $v \sin i = 6.9 \pm
0.8$~km~s$^{-1}$ (\cite{dfp+98}); this rotation velocity can be
reconciled with the observed starspot modulation period only if the
inclination is high ($\ge 60$~deg). The rotational velocity of M
dwarfs (\cite{dfp+98}) is characterized by the bulk of them having a
narrow distribution with $v \sin i \le 5$~km~s$^{-1}$, and a tail of
rare fast rotators with velocities of up to $\simeq 50$~km~s$^{-1}$.
EV~Lac lies at the upper end of the slow-rotator distribution.

The absolute magnitude is $M_V = 11.73$, which, given a color index
$(R-I)_{\rm C} = 1.52$ translates in $M_{\rm bol} = 9.40$
(\cite{dfp+98}), corresponding to $L_{\rm bol} = 5.25 \times
10^{31}$~erg~s$^{-1}$. Using the mass-luminosity relationship of
\cite*{bca+98} and the $K$-band luminosity (to be preferred given the
independence of the relationship between mass and $K$-band luminosity
from metallicity) $M_K = 6.78$ (\cite{dfp+98}) we derive a mass of
$\simeq 0.35~M_\odot$. No photospheric abundance analyses of EV~Lac
are known to us, although \cite*{fsg95} report a near-solar
metallicity based on broad-band photometry. The radius for a
solar-metallicity $0.35~M_\odot$ dwarf is, from the models of
\cite*{cb97}, $R \simeq 2.5 \times 10^{10}$~cm, or $R \simeq
0.36~R_\odot$, (assuming the star is old enough, given that such a
low-mass star will contract gravitationally until it's $\simeq
300$~Myr old).

Stellar structure models show that the radiative core of low-mass
stars shrinks with decreasing mass, disappearing completely in mid-M
dwarfs, so that late-M dwarfs are expected to be completely
convective. Solar-type dynamos are thought to require an interface
between the convective envelope and the radiative core (the
$\alpha$--$\Omega$ shell dynamo model) and are thus not expected to be
present in the cooler M dwarfs. Given that however X-ray activity is
present and common even in purportedly fully convective late M dwarfs
(\cite{bms+93}; \cite{sfg95}) a different type of dynamo mechanism
must be operating in them; it has recently been suggested that
small-scale magnetic fields can be generated in convection zones by a
turbulently driven dynamo (\cite{dyr93}). This -- which could be also
at work in higher mass stars with varying degrees of efficiency,
depending on the rotation rate -- would therefore provide the only
magnetic field generation mechanism in fully convective low-mass
dwarfs.

The predicted mass at which stars become fully convective depends on
the details of the physics adopted in the stellar models.
\cite*{cb97} use non-grey atmospheres to put the fully convective
limit at $0.35~M_\odot$ (i.e.\ just at the estimated mass of EV~Lac),
independent of metallicity in the range $0.01\times Z_\odot < Z <
Z_\odot$.  Different calibrations for the mass-luminosity
relationships (see discussion in \cite{dfp+98}) push the
fully-convective limit toward somewhat higher masses, so that EV~Lac
is most likely fully convective, and thus its activity is likely to be
driven purely by a turbulent dynamo, which is not, among other things,
expected to have a solar-like cyclic behavior, and which is expected
to generate magnetic fields on a smaller spatial scale (related to the
scale of the turbulent flow fields) than an $\alpha$--$\Omega$ dynamo.
The picture is complicated by the fact that strong magnetic fields may
influence the stellar interior structure maintaining a radiative core
at masses lower than the theoretical limits for spherically symmetric,
non-magnetic stars (\cite{csh81}), and thus the magnetic fields of
very active, low-mass stars could still be (partially) powered by an
$\alpha$--$\Omega$ shell dynamo.

\subsection{Previous X-ray observations}

The high activity of EV~Lac had been noted well before the advent of
high-energy observations from its large optical and UV flaring rate,
with some truly exceptional optical flares observed: \cite*{rs82}
report the occurrence of a 6.4~mag $U$-band flare, lasting 6.4~hr,
with a peak luminosity of $\simeq 10^{32}$~erg~s$^{-1}$ and a total
energy output of $\ge 10^{35}$~erg.

EV~Lac was first observed as a quiescent soft X-ray source by EXOSAT
(\cite{sr85}) -- although it had been detected earlier by HEAO-1
during a strong flare with $\log L_{\rm X} = 28.7$~erg~s$^{-1}$. It
was not observed by the \emph{Einstein} observatory, while it was
observed by ROSAT both in the All Sky Survey (RASS -- in which a major
flare was also detected) and in pointed mode.  The RASS quiescent
X-ray luminosity was $\log L_{\rm X} = 29.08$~erg~s$^{-1}$
(\cite{sfg95}), corresponding to $\log L_{\rm X}/L_{\rm bol} = -2.6$.
It was also the subject of one SAX and several PSPC pointings,
analyzed in \cite*{smf+99}, during which its coronal emission shows
both a continuous variability of about a factor 2--3 and the
occurrence of an intense flare, with an increase of the emission in
the PSPC of about 10-fold.

The HEAO A-1 Sky Survey experiment (\cite{asw84}) detected two flares
from EV~Lac, with peak X-ray luminosities (in the 0.5--20~keV band) of
$\log L_{\rm X} = 29.5$~erg~s$^{-1}$ and $\log L_{\rm X} =
30.3$~erg~s$^{-1}$. The most energetic of the two flares represented a
peak enhancement of $\simeq 50$ over the quiescent X-ray luminosity
($L_{\rm X} \simeq 10^{28.5}$~erg~s$^{-1}$), and its decay $e$-folding
time was roughly estimated to be of the order of $\le 5$~ks (although
the very sparse temporal coverage prevents an accurate determination
of the flare's decay). \cite*{asw84} have also estimated the physical
parameters of the two flares (\emph{assuming} a peak temperature of
$\simeq 2 \times 10^7$~K), resulting, for the smaller of the two
flares, in a peak emission measure $EM \simeq 2 \times
10^{53}$~cm$^{-3}$, a density $n \simeq 5 \times 10^{11}$~cm$^{-3}$
and a loop length $L \simeq 5 \times 10^{9}$~cm.  The magnetic field
strength necessary to confine the plasma was estimated at $200~G$.
Only the peak emission measure is reported for the second flare ($EM
\simeq 2 \times 10^{53}$~cm$^{-3}$). A small flare was seen in one of
the EXOSAT observations, in the LE detector, as discussed in detail by
\cite*{pts90}. Its rise time ($1/e$ time) was $\simeq 600$~s, and its
decay time was $\simeq 4.5$~ks. At peak the flare represented an
enhancement of only $\simeq 3$ times over the quiescent X-ray flux,
with a peak X-ray luminosity $L_{\rm X} \simeq 10^{29}$~erg~s$^{-1}$
and a total energy released in the X-rays $E \simeq 10^{32}$~erg. The
lack of energy resolution of the EXOSAT LE detector made it impossible
to perform an analysis of the flare's decay.

\cite*{sch94} derived loop parameters for the long RASS flare by
fitting the flare decay parameters within the framework of the
quasi-static formalism of \cite*{om89}. The maximum observed
temperature is $T \simeq 30$~MK, the decay timescale is $\tau \simeq
38$~ks, and the peak emission measure is $EM \simeq 1.5 \times
10^{52}$~cm$^{-3}$. The loop length derived through a quasi-static
analysis is large, at $ L \simeq 6 \times 10^{11}$~cm~$\simeq 10~R_*$
(with an inferred flaring volume $V \simeq 3 \times 10^{31}$~cm$^3$)
and the plasma density is correspondingly low, at $n \simeq 3 \times
10^{10}$~cm$^{-3}$. The total thermal energy was estimated at $E
\simeq 9 \times 10^{34}$~erg.  \cite*{sch94} also analyzed the EV~Lac
PSPC flare decay using the two-ribbon model of \cite*{kp84}.

\end{document}